\documentclass{article}
\usepackage[T1]{fontenc}

\usepackage[utf8]{inputenc}
\usepackage{arxiv_ismir} 
\usepackage{amsmath,cite,url}
\usepackage{graphicx}
\usepackage{color}
\usepackage{booktabs}
\usepackage{tabularx}
\usepackage{array,makecell}
\usepackage{amssymb}
\usepackage{xcolor}
\definecolor{taskred}{HTML}{B85450}
\definecolor{taskblue}{HTML}{6C8EBF}
\definecolor{taskgreen}{HTML}{82B366}

\title{{Gangqinpu}: A Multi-Genre, Multi-Task Piano Dataset with Aligned Audio, MIDI, and Musical Scores}

\multauthor
  {Wei Zeng$^{1,2}$ \hspace{1cm} Anvay Mathur$^3$ \hspace{1cm} Ye Wang$^{1,2}$}
  {
  $^1$ Integrative Sciences and Engineering Programme, NUS Graduate School\\
  $^2$ School of Computing, National University of Singapore\\
  $^3$ Singapore American School\\
  {\tt\small w.zeng@u.nus.edu, mathur47349@sas.edu.sg, dcswangy@nus.edu.sg}
  }

\begin{document}
\pagestyle{plain}
\maketitle

\begin{abstract}
Aligned audio, performance MIDI, and musical scores provide valuable supervision for music analysis and performance modeling, yet datasets that jointly align all three modalities remain scarce and limited in scale and genre diversity. We introduce \textbf{Gangqinpu}, a multi-genre piano dataset containing 18,786 score--performance pairs and 1,007.1 hours of source audio, with automatic note-level alignments and derived beat/downbeat annotations. It spans diverse genres, including pop, classical, jazz, folk/traditional, rock, and blues. Our automated pipeline transcribes piano audio, aligns the resulting MIDI with musical scores, assigns genre labels, and filters out non-expressive performances with a score-informed binary classifier. We further construct joint splits with ASAP using conservative score-similarity grouping to reduce data leakage. We evaluate Gangqinpu on performance MIDI-to-score transcription, expressive performance rendering, and beat tracking. Downstream experiments demonstrate the value of Gangqinpu as complementary training data, with cross-dataset gains in beat tracking and primarily in-domain benefits for transcription and expressive performance rendering. The code and dataset are publicly available at \url{https://github.com/wei-zeng98/gangqinpu}.
\end{abstract}

\section{Introduction}

Piano music has long been an important research domain in music information retrieval (MIR), owing to its extensive and stylistically diverse repertoire, rich polyphonic structure, and availability in both acoustic and symbolic representations. A piano piece and its performance can be represented through three complementary modalities. \textit{Audio} captures the acoustic realization, \textit{performance MIDI} describes note-level performance details, and a \textit{musical score} encodes the underlying compositional structure.

\begin{figure}[t]
\centering
\includegraphics[width=\linewidth]{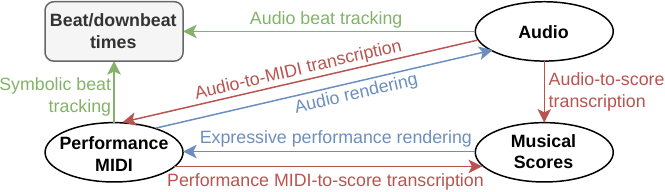}
\caption{Musical representations and associated tasks. Arrows indicate task input--output relationships; colors distinguish transcription (\textcolor{taskred}{red}), rendering (\textcolor{taskblue}{blue}), and beat tracking (\textcolor{taskgreen}{green}).}
\label{fig:representations_tasks}
\end{figure}

\autoref{fig:representations_tasks} illustrates these relationships. Transcription converts audio into performance MIDI~\cite{yan2021skipping,yan2024scoring,kong2021high,hawthorne2021sequence,toyama2023automatic,hawthorne2018onsets,jung2026umust} or musical scores~\cite{zeng2024end,carvalho2017towards,roman2019holistic,tamer2026rubato,hu2026tutti,wang2026learned}. It can also recover scores from performance MIDI~\cite{liu2022performance,beyer2024end,zeng2026bridging}. Expressive performance rendering predicts timing and dynamics from scores~\cite{zeng2026bridging,fujishima2019rendering,jeong2019virtuosonet,rhyu2022sketching,borovik2023scoreperformer,renault2023expressive,tang2023reconstructing,zhang2024dexter,tang2025towards}, while beat tracking estimates beat and downbeat times from audio or performance MIDI~\cite{zhao2022beat,foscarin2024beat,he2026joint,murgul2025beat}. Aligned datasets with rhythmic annotations thus support multiple tasks in music analysis and performance modeling.

Despite the importance of data connecting these modalities, existing piano datasets typically provide only a subset of audio, performance MIDI, and musical scores. Datasets that jointly align all three modalities remain scarce. ASAP~\cite{foscarin2020asap} is a notable example, but it provides audio for only 519 of its 1,067 MIDI performances, with a small number of performances not fully aligned to their corresponding scores. Moreover, its repertoire is restricted to Western classical piano music. The broader predominance of classical repertoire in existing datasets limits their coverage of musical styles and applications.

Beyond classical music, pop and jazz performances may involve lead sheets, informal arrangements, or improvisation rather than fully notated scores, making transcription from recordings particularly useful. Expressive performance rendering also has potential applications in contemporary music production by transforming quantized MIDI into natural-sounding performances within digital audio workstations (DAWs). However, research on expressive performance rendering has focused predominantly on classical piano \cite{bontempi2023research}. These applications motivate larger aligned piano datasets spanning a broader range of genres.

To address this gap, we introduce \textbf{Gangqinpu}, a multi-genre piano dataset with aligned audio, transcribed performance MIDI, and musical scores. We construct the dataset from online audio--score pairs using an automated pipeline for audio transcription, expressive performance filtering, genre annotation, and note-level score alignment. A score-informed binary classifier filters out non-expressive performances with limited timing and dynamic variation, including renderings synthesized directly from scores. The dataset spans diverse genres, including pop, classical, jazz, folk/traditional, rock, and blues, with additional samples labeled as other or unknown.

The main contributions of this work are as follows:

\begin{itemize}
    \item We introduce \textbf{Gangqinpu}, a multi-genre piano dataset with 18,786 score--performance pairs, 1,007.1 hours of source audio, and automatic note-level and beat/downbeat annotations.

    \item We develop an automated \textbf{construction pipeline} for transcription, alignment, genre annotation, and expressive performance filtering, with joint score grouping across Gangqinpu and ASAP to reduce data leakage.

    \item We demonstrate Gangqinpu's value as complementary training data for beat tracking, performance MIDI-to-score transcription, and expressive performance rendering. Gains extend across datasets for beat tracking and are primarily in-domain for transcription and rendering.
\end{itemize}

\section{Related Work}
\subsection{Piano Datasets}

Piano performance MIDI datasets are typically constructed through direct capture or automatic music transcription (AMT). MAESTRO~\cite{hawthorne2019maestro} provides synchronized audio and MIDI captured using Yamaha Disklavier pianos, but this approach requires specialized recording equipment. AMT enables more scalable collection from existing recordings: ATEPP~\cite{zhang2022atepp} provides transcribed performances by professional pianists, while Aria-MIDI~\cite{bradshaw2025aria} scales up to over one million MIDI files from online recordings, both subject to transcription errors.

For tasks that relate performances to musical notation, performance MIDI must be paired with corresponding scores. ASAP~\cite{foscarin2020asap} provides beat-level alignments, and (n)ASAP~\cite{peter2023automatic} extends these to the note level through automatic alignment and manual inspection. ACPAS~\cite{liu2021acpas} combines real and synthesized audio with MIDI scores and beat-level annotations. ATEPP provides corresponding MusicXML scores for a subset of its recordings, while PianoCoRe~\cite{borovik2026pianocore} consolidates existing corpora into a large note-aligned MIDI collection.

\autoref{tab:dataset_comparison} summarizes these datasets in terms of scale, modalities, and annotations. Existing score-aligned resources predominantly cover Western classical repertoire, where performances typically follow established scores. Extending such resources to popular music is challenging because recordings may be produced without a written score~\cite{bontempi2023research}. POP909~\cite{wang2020pop909} supports popular-music research through MIDI arrangements with beat, chord, and key annotations, but does not provide piano recordings paired with independently notated scores. Gangqinpu addresses this gap by connecting multi-genre piano recordings with musical scores through transcribed performance MIDI and automatic note-level alignments.

\subsection{Audio-to-MIDI Piano Transcription and Score--Performance Alignment}

Recent piano transcription systems include autoregressive encoder--decoder models such as Aria-AMT~\cite{bradshaw_aria_amt} and interval-based models such as Transkun v2~\cite{yan2024scoring}. Aria-AMT predicts note and pedal tokens, whereas Transkun v2 combines Transformer-based interval scoring with non-autoregressive semi-CRF decoding. We select Transkun v2 for its structured prediction of note and pedal intervals and its ability to preserve note offsets without pedal-based extension, allowing note offsets and sustain-pedal activity to be represented separately.

For symbolic alignment, Nakamura et al.~\cite{nakamura2017performance} combine HMM-based matching with error detection and local realignment using merged-output HMMs. DualDTWNoteMatcher~\cite{peter2023online}, implemented in Parangonar, instead separates pitch-sequence matching from pitch-wise onset alignment. Across four evaluated piano datasets, it achieved a higher combined note-matching F1 than the Nakamura baseline (99.0\% versus 98.5\%). These reported results and its integration with our symbolic processing and refinement pipeline motivate our choice of DualDTWNoteMatcher.

\begin{table*}[t]
\centering
\scriptsize
\setlength{\tabcolsep}{4pt}
\renewcommand{\arraystretch}{1.15}
\begin{tabular}{@{}llrrrcccccc@{}}
\toprule
& &
\multicolumn{3}{c}{\textbf{Size}} &
\multicolumn{3}{c}{\textbf{Modalities}} &
\textbf{Origin} &
\multicolumn{2}{c}{\textbf{Annotations}} \\
\cmidrule(lr){3-5}
\cmidrule(lr){6-8}
\cmidrule(lr){10-11}
\textbf{Dataset} &
\textbf{Genre} &
\textbf{Pieces / Scores} &
\textbf{Perf.} &
\textbf{Hours} &
\textbf{Audio} &
\textbf{MIDI} &
\textbf{Score} &
\textbf{Audio / MIDI Source} &
\textbf{Align.} &
\textbf{Beats} \\
\midrule

MAESTRO v3~\cite{hawthorne2019maestro}
& Mostly classical
& --
& 1,276
& 198.7
& $\checkmark$
& $\checkmark$
& --
& Real / Captured
& --
& --
\\

ASAP/(n)ASAP~\cite{foscarin2020asap,peter2023automatic}
& Classical
& 235
& 1,067
& 94.5
& External$^{\triangle}$
& $\checkmark$
& $\checkmark$
& Real / Captured
& Beat / Note$^\dagger$
& $\checkmark$
\\

ATEPP v1.2~\cite{zhang2022atepp}
& Classical
& 1,595
& 11,674
& $\sim$1,000
& External
& $\checkmark$
& $\triangle^\ast$
& Real (web) / AMT
& --
& --
\\

ACPAS~\cite{liu2021acpas}
& Classical
& 497
& 2,189
& 179.8
& Ext./Synth.
& $\checkmark$
& $\checkmark$
& Real+Synth. / Mixed
& Beat$^\S$
& $\checkmark$
\\

PianoCoRe-A~\cite{borovik2026pianocore}
& Primarily classical
& 1,591
& 157,207
& 12,509
& --
& $\checkmark$
& $\checkmark$
& None / Mixed
& Note
& --
\\

POP909~\cite{wang2020pop909}
& Pop
& 909
& --
& $\sim$60
& --
& $\checkmark$
& --
& Reference / Arrangement
& --
& $\checkmark$
\\

\midrule
\textbf{Gangqinpu} & \textbf{Multiple} & \textbf{15,645} & \textbf{18,786} & \textbf{1,007.1} & \textbf{External} & $\checkmark$ & $\checkmark$ & \textbf{Web / AMT} & \textbf{Note} & $\checkmark$
\\

\bottomrule
\end{tabular}

\vspace{3pt}
\begin{minipage}{0.99\textwidth}
\footnotesize
$\checkmark$ indicates availability, $\triangle$ partial coverage, and -- unavailable or unreported information. External (Ext.) audio is obtained separately; Gangqinpu provides audio URLs and a download script. Align.\ denotes score--performance alignment. Captured denotes instrument-recorded MIDI, AMT automatic transcription, and Web online audio of unverified recording/synthesis origin. $^\dagger$ Statistics follow (n)ASAP~\cite{peter2023automatic}; 1,062 performances have note alignments. $^\ast$ MusicXML covers roughly half the tracks. $^\S$ 30 performances lack alignment.
\end{minipage}

\caption{Comparison of piano datasets by scale, modalities, data origin, and score--performance annotations.}
\label{tab:dataset_comparison}
\end{table*}

\section{Dataset Construction}
\subsection{Overall Pipeline}

\begin{figure}[t]
    \centering
    \includegraphics[width=\linewidth]{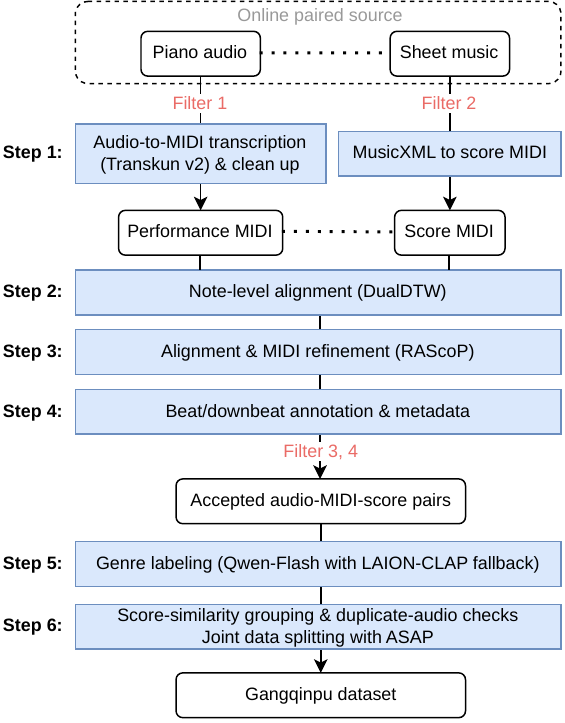}
    \par\vspace{3pt}
    \begin{minipage}{\linewidth}
        \footnotesize
        Dotted horizontal links indicate paired representations. Filters retain samples satisfying: \textbf{(1)} audio duration $\geq 60$\,s; \textbf{(2)} at least 32 measures and piano-only parts; \textbf{(3)} insertion and deletion rates each $\leq 10\%$, measured from the raw alignment before refinement; \textbf{(4)} predicted expressive-performance probability $\geq 0.5$.
    \end{minipage}
    \caption{Overview of the Gangqinpu dataset construction pipeline.}
    \label{fig:pipeline}
\end{figure}

Figure~\ref{fig:pipeline} summarizes the construction of Gangqinpu from online audio--score pairs. After filtering by audio duration, score length, and instrumentation, we transcribe audio into performance MIDI using Transkun v2~\cite{yan2024scoring}, clean the resulting MIDI, and convert MusicXML scores into score MIDI. DualDTWNoteMatcher~\cite{peter2023online} establishes note-level correspondences between the two representations, followed by alignment and MIDI refinement using RAScoP~\cite{borovik2026pianocore}. We then derive beat/downbeat annotations and metadata, retaining samples that satisfy raw alignment error thresholds and are classified as expressive performances (see Section \ref{sec:filtering}). Accepted samples receive genre labels through Qwen-Flash~\cite{alibaba_qwen_flash}, with LAION-CLAP~\cite{wu2023clap} as a fallback. Finally, score-similarity grouping and duplicate-audio checks guide joint data splitting with ASAP to reduce leakage across partitions and datasets.

\subsection{Data Acquisition and Preprocessing}

We collect publicly accessible piano audio and associated sheet music from online sources, together with metadata such as titles and contributor information. We retain recordings lasting at least 60 seconds (Filter 1) and scores containing at least 32 measures with piano-only parts, excluding short excerpts and scores for other instrumentation (Filter 2).

Scores are converted into MusicXML and processed using the SyMuPe and Partitura toolkits~\cite{borovik2025symupe,cancinochacon2022partitura}. Repeats and alternative endings are unfolded, and a score MIDI representation is generated while retaining links to score positions and metrical information. Before alignment, performance tracks are merged into a single note sequence, duplicate and implausibly short notes are removed, and overlapping notes of the same pitch are truncated. The resulting representations retain score structure and expressive performance information for note-level alignment and annotation.

\subsection{Alignment-Derived Beat Annotation and Confidence}

We obtain initial note correspondences using Parangonar's DualDTWNoteMatcher~\cite{peter2023online} and refine them with RAScoP~\cite{borovik2026pianocore}. Interpolated performance notes introduced during refinement are explicitly marked to distinguish them from observed transcription events. To assess score--performance consistency, we compute insertion and deletion rates from the raw alignment, normalized by the numbers of performance and score notes, respectively. Pairs are excluded if either rate exceeds 10\% (Filter 3), preventing subsequent interpolation from masking substantial mismatches.

A metrical grid is extracted from the unfolded score, preserving time-signature changes and using dotted-quarter-note beats in compound meters such as 6/8, 9/8, and 12/8. Score beats are transferred to performance time through the refined note correspondences. Inspired by the ASAP annotation procedure~\cite{foscarin2020asap}, we identify notes within $\pm0.05$ quarter notes of each score beat and use the median of their corresponding performance onset times as the beat time. Beats without direct support are linearly interpolated between neighboring onset anchors or extrapolated at the boundaries using a local tempo estimate. Downbeats and time/key-signature changes are mapped to the same performance timeline.

We assign heuristic confidence scores based on the evidence supporting each beat. A beat initially receives a confidence of 0.95 if any nearby mapped note is observed, or 0.40 if all are interpolated. When beat times are interpolated between neighboring onset anchors, confidence is 0.75 times the lower anchor confidence; extrapolated beats receive 0.25. If any score note within $\pm0.05$ quarter notes of the beat is marked as interpolated, confidence is multiplied by 0.35. We mark beats with confidence below 0.50 as low confidence and those below 0.35 as ambiguous. These scores reflect heuristic annotation reliability rather than calibrated probabilities and support masking uncertain regions in downstream tasks.

\subsection{Expressive Performance Filtering}
\label{sec:filtering}
Online audio--score collections often include deadpan renderings synthesized directly from notation, which sound mechanical due to rigid timing and little or no expressive variation in dynamics. Although useful for some transcription tasks, such examples provide limited supervision for expressive performance rendering, where the objective is to learn musically meaningful deviations from the score. We therefore distinguish expressive from non-expressive performances based on perceived timing and dynamics, regardless of production origin. Human recordings and digitally edited MIDI renderings are both eligible if they exhibit sufficient expressiveness.

To automate this selection, we train a binary classifier that takes a performance MIDI and its corresponding score as input (Filter 4). We construct a labeled subset of Gangqinpu through independent listening assessments by two human experts. Each expert assigns an expressive or non-expressive label according to the audible performance. Disagreements are discussed and reassessed through repeated listening until consensus is reached. The resulting subset contains 779 expressive and 370 non-expressive examples.

The classifier is a multilayer perceptron (MLP) operating on eight handcrafted features summarized in \autoref{tab:expressiveness_features}. Features are extracted from observed note correspondences in the raw alignment, avoiding interpolated events introduced during refinement. To measure timing variation, we first group notes with the same score onset and represent each group by its median performed onset. At each onset position, we fit a local linear mapping from score time to performance time within a window of $\pm4$ quarter notes, widening it when necessary to include at least six distinct onset positions, where available. Differences between observed and predicted onset times yield three features: timing variability, correlation, and entropy. Separate features measure onset differences within chords and variation in note velocities.

The standardized feature vector is processed by two hidden layers with 32 and 16 units and ReLU activations, followed by a binary probabilistic output. Training uses Adam, $L_2$ regularization with $\alpha=10^{-3}$, and early stopping. The labeled subset is supplemented with 982 usable ASAP performances as positive examples and synthetically perturbed score renderings as negative examples. Each source is partitioned into training and held-out subsets at a 75:25 ratio, with synthetic variants grouped by their underlying score; class imbalance is addressed by oversampling within the training set. Samples with a predicted expressive-class probability of at least 0.5 are retained.

\begin{table}[t]
\centering
\small
\setlength{\tabcolsep}{4pt}
\renewcommand{\arraystretch}{1.12}
\begin{tabularx}{\linewidth}{@{}lX@{}}
\toprule
\textbf{Feature} & \textbf{Definition} \\
\midrule
Timing variability & Standard deviation of local onset residuals. \\
Timing correlation & Lag-one autocorrelation of onset residuals. \\
Timing entropy & Shannon entropy of onset residuals discretized into 10 ms bins. \\
Chord asynchrony & Median onset range within groups of simultaneously notated notes. \\
Velocity variability & Standard deviation of note velocities. \\
Velocity diversity & Number of distinct note velocities. \\
Staff contrast & Mean velocity of upper-staff notes minus that of notes on other staves. \\
Pitch contrast & Mean within-chord velocity difference between the highest note and the remaining notes. \\
\bottomrule
\end{tabularx}
\caption{Score-informed timing and dynamic features used for expressive performance classification.}
\label{tab:expressiveness_features}
\end{table}

\subsection{Genre Annotation and Leakage-Aware Splitting}

We assign sample-level genre labels using a two-stage procedure. Qwen-Flash~\cite{alibaba_qwen_flash} first classifies each sample from its textual metadata, including the title and contributor information. Samples whose genre remains undetermined are subsequently assessed using LAION-CLAP~\cite{wu2023clap} with the music checkpoint, which compares audio embeddings with textual genre descriptions. Predictions that do not meet the confidence criteria remain labeled as unknown. Labels are normalized into six named genres (pop, classical, jazz, folk/traditional, rock, and blues), along with the additional categories of other and unknown.

To reduce leakage, we identify related scores before assigning samples to training, validation, and test sets. Candidate pairs are retrieved using transposition-invariant pitch-interval patterns from the upper and lower voices, then verified through monotonic alignment and polyphonic note matching. We maintain two grouping levels in the metadata. Score-similarity groups (\texttt{score\_group\_id}) are used to report the number of score groups in the dataset. Broader split groups (\texttt{group\_id}) additionally link potentially related arrangements and excerpts, including matches between Gangqinpu and ASAP, and are used only for partition assignment. Each split group is assigned entirely to one partition, retaining multiple performances of the same score together. Exact audio duplicates are also checked across partitions.

We jointly group Gangqinpu and (n)ASAP v2.1.1~\cite{foscarin2020asap,peter2023automatic} to account for cross-dataset overlap. Groups containing the ASAP test pieces used by \cite{beyer2024end} are assigned to the test set. The remaining groups are assigned to training, validation, or test, with each group kept in a single partition. The resulting train/validation/test splits contain 17,773/499/514 Gangqinpu samples and 955/50/61 (n)ASAP samples. All downstream tasks in Section~\ref{sec:experiments} use these fixed assignments, selecting samples with the required inputs and annotations. Evaluation is reported separately on the two datasets.

\section{Dataset Statistics}
\label{sec:dataset_statistics}

\begin{figure*}[t]
\centering
\includegraphics[width=\linewidth]{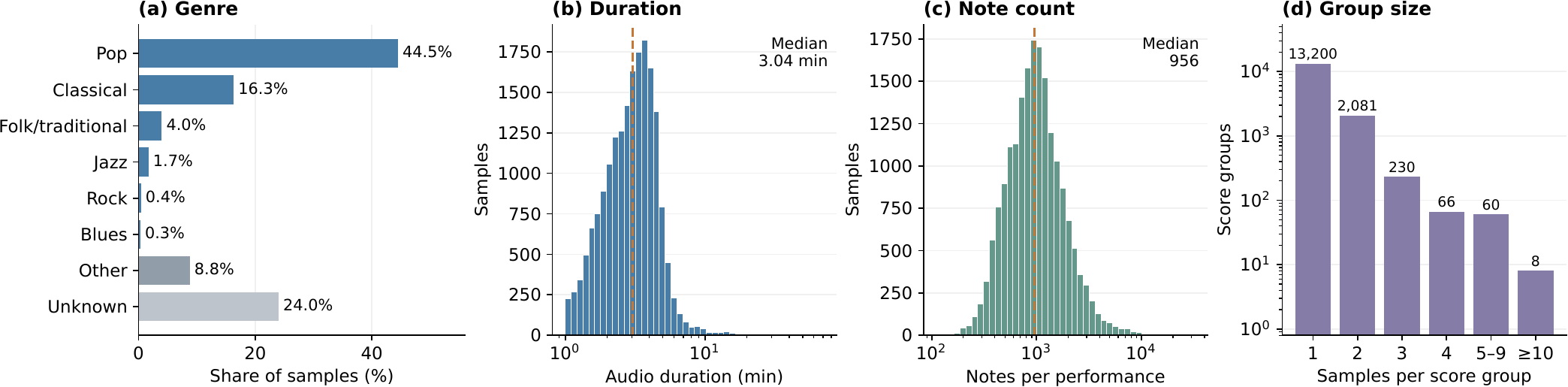}
\caption{Distributions of (a) automatically assigned genre labels, (b) audio duration, (c) refined performance MIDI note counts, including interpolated notes, and (d) performances per score group. Panels (b) and (c) use logarithmic horizontal axes; panel (d) uses a logarithmic vertical axis. Dashed lines indicate medians.}
\label{fig:gangqinpu_distributions}
\end{figure*}

Gangqinpu contains 18,786 paired performances totaling 1,007.1 hours of source audio, organized into 15,645 automatically identified score groups. These groups capture identical or highly similar scores, including transposed variants, and serve as an operational measure of score diversity rather than a verified count of distinct compositions. Among them, 2,445 groups contain multiple performances, providing alternative realizations of similar musical material. \autoref{tab:gangqinpu_statistics} summarizes the dataset scale, and \autoref{fig:gangqinpu_distributions} presents its principal distributions.

\autoref{fig:gangqinpu_distributions} summarizes the genre, duration, and note-count distributions. Pop and classical dominate the automatically assigned genre labels, alongside folk/traditional, jazz, rock, and blues; uncertain assignments remain unknown. Audio duration and note counts exhibit long-tailed distributions, reflecting variation in recording length and sequence size. Note counts are computed from refined performance MIDI and include interpolated notes.

\begin{table}[t]
\centering
\small
\setlength{\tabcolsep}{5pt}
\begin{tabular}{@{}lr@{}}
\toprule
\textbf{Statistic} & \textbf{Value} \\
\midrule
Performances & 18,786 \\
Score groups & 15,645 \\
Groups with multiple performances & 2,445 \\
Total audio duration (hours) & 1,007.1 \\
Mean audio duration (min) & 3.22 \\
Median audio duration (min) & 3.04 \\
Median notes per performance & 956 \\
\bottomrule
\end{tabular}
\caption{Summary statistics of Gangqinpu. Score groups are identified automatically through score-similarity analysis.}
\label{tab:gangqinpu_statistics}
\end{table}

\section{Experiments}
\label{sec:experiments}

\subsection{Expressive Performance Classification}

We evaluate whether the proposed classifier distinguishes expressive from non-expressive performances according to expert listening judgments, irrespective of recording or production provenance, using four subsets: \textit{ASAP}, real human performances; \textit{expressive} and \textit{non-expressive}, the two classes of our own expert-labeled subset; and \textit{hard\_neg}, a synthetic negative class rendered directly from scores with small Gaussian onset/velocity noise ($\sigma_\text{onset}\!\leq\!30$\,ms, $\sigma_\text{velocity}\!\leq\!9$). After alignment and feature-validity checks, the expert-labeled subset contains 779 expressive and 370 non-expressive samples; a fixed 75/25 split within each class yields 195 expressive and 93 non-expressive held-out samples, with ASAP and hard\_neg likewise split 75/25 to supplement training and evaluation.

The eight-feature MLP achieves 98.61\% accuracy and 98.41\% macro-F1 on the 288 expert-labeled held-out samples (Table~4). It correctly identifies 193 of 195 expressive performances and 91 of 93 non-expressive performances, with two errors in each class. These results support its use as an automatic filtering component for the collected data, although they do not establish generalization to other transcription pipelines or data sources.

\textbf{Comparison with existing classifiers.} We compare against two published, readily available classifiers addressing related tasks: the SyMuPe quality classifier~\cite{borovik2025symupe,borovik2026pianocore}, a four-way MIDI quality model (\textit{score}, \textit{high quality}, \textit{low quality}, \textit{corrupted}), mapped to binary here by treating \textit{high quality} as expressive and the remaining three categories as non-expressive; and NOMML, the median-metric-level measure from GigaMIDI~\cite{lee2025gigamidi}, applied with its reported threshold (median level $=12$). Evaluation uses the same 534 held-out samples as Table~4, plus 480 synthetic samples (120 each) from the hard\_neg and 46/60/80\,ms noise conditions.

\begin{table}[t]
\centering
\resizebox{\columnwidth}{!}{%
\begin{tabular}{lrrrr}
\toprule
Subset & $n$ & Ours & SyMuPe & NOMML \\
\midrule
ASAP (human)               & 246 & 100.0\% & 99.6\%  & 100.0\% \\
expressive         & 195 & 99.0\%  & 86.7\%  & 100.0\% \\
non-expressive    & 93  & 97.8\%  & 18.3\%  & 0.0\%   \\
hard\_neg (synthetic)        & 120 & 100.0\% & 100.0\% & 0.0\%   \\
noise 46\,ms (synthetic)     & 120 & 98.3\%  & 100.0\% & 0.0\%   \\
noise 60\,ms (synthetic)     & 120 & 83.3\%  & 100.0\% & 0.0\%   \\
noise 80\,ms (synthetic)     & 120 & 30.8\%  & 100.0\% & 0.0\%   \\
\midrule
Pooled accuracy ($n{=}534$) &     & \textbf{99.25\%} & 80.71\% & 82.58\% \\
Pooled macro-F1             &     & \textbf{98.70\%} & 56.88\% & 45.23\% \\
\bottomrule
\end{tabular}%
}
\caption{Comparison with SyMuPe and NOMML. Per-subset values are detection rates for that subset's single ground-truth class; pooled rows combine the three real-data splits ($n{=}534$), the only ones mixing both classes.}
\label{tab:baseline}
\end{table}

\autoref{tab:baseline} shows that our classifier achieves the highest pooled accuracy and macro-F1 on the real-data evaluation subset, while its detection rate decreases as synthetic onset noise increases. SyMuPe recalls expressive performances well but labels 76 of 93 non-expressive samples as \textit{high quality}, reflecting its design for general quality triage rather than score-synchronized expressiveness detection. NOMML assigns a metric level of 12 to every sample and therefore predicts all samples as expressive. Its pooled accuracy of 82.58\% reflects the class distribution (441 expressive and 93 non-expressive), while its macro-F1 is 45.23\%. These comparisons assess the classifiers' suitability for our filtering task rather than their general performance.

\subsection{Downstream Task Evaluation}
\label{sec:downstream_evaluation}

We evaluate the utility of Gangqinpu on three complementary tasks: beat tracking, performance MIDI-to-score transcription, and expressive performance rendering. We use Beat This!~\cite{foscarin2024beat} for estimating beat and downbeat positions from audio, PM2S~\cite{liu2022performance} for recovering score-related symbolic attributes from performance MIDI, and an adapted S2A~\cite{tang2025towards} model for predicting expressive timing and velocity from a musical score. These tasks assess the usefulness of the dataset's rhythmic annotations, score--performance correspondences, and expressive performance representations, respectively.

For each baseline, we compare training on ASAP alone with training on ASAP combined with Gangqinpu and evaluate both conditions separately on the ASAP and Gangqinpu test sets. All tasks follow the fixed joint split assignments, with usable subsets determined by the required inputs and annotations. The ASAP test subsets contain 26 samples for Beat This!, 59 for PM2S hand assignment, and 61 for the remaining PM2S tasks and S2A. All tasks use 514 Gangqinpu test samples.

\subsubsection{Beat Tracking}
\label{sec:beat_tracking}
We evaluate Beat This!~\cite{foscarin2024beat} on beat and downbeat tracking using F1, CMLt, and AMLt~\cite{davies2009evaluation}. Low-confidence and temporally inconsistent annotation regions, together with regions outside the annotated span, are masked during training and evaluation. Both predicted and reference events within masked regions are excluded from metric computation. \autoref{tab:beat_tracking} compares the two training conditions on both test sets. Adding Gangqinpu improves all reported metrics on both datasets. On ASAP, beat and downbeat F1 increase by 4.30 and 4.31 percentage points, respectively. The gains are larger on Gangqinpu, where beat F1 improves from 60.18 to 78.16 and downbeat F1 from 39.60 to 63.66. Higher CMLt and AMLt scores indicate more reliable beat-sequence tracking. These results suggest that Gangqinpu provides useful rhythmic supervision, with benefits extending to the ASAP test set.

\begin{table}[t]
\centering
\small
\setlength{\tabcolsep}{4pt}
\begin{tabular}{@{}llrrr@{}}
\toprule
\textbf{Target} & \textbf{Training data} & \textbf{F1} $\uparrow$  & \textbf{CMLt} $\uparrow$ & \textbf{AMLt} $\uparrow$ \\
\midrule
\multicolumn{5}{c}{\textit{ASAP test}} \\
\midrule
Beat & ASAP & 57.69 & 21.66 & 26.63 \\
     & ASAP + Gangqinpu & \textbf{61.98} & \textbf{28.57} & \textbf{30.64} \\
Downbeat & ASAP & 43.23 & 10.90 & 21.30 \\
         & ASAP + Gangqinpu & \textbf{47.55} & \textbf{13.41} & \textbf{22.79} \\
\midrule
\multicolumn{5}{c}{\textit{Gangqinpu test}} \\
\midrule
Beat & ASAP & 60.18 & 28.64 & 35.09 \\
     & ASAP + Gangqinpu & \textbf{78.16} & \textbf{58.42} & \textbf{63.63} \\
Downbeat & ASAP & 39.60 & 9.26 & 19.91 \\
         & ASAP + Gangqinpu & \textbf{63.66} & \textbf{30.51} & \textbf{51.63} \\
\bottomrule
\end{tabular}
\caption{Beat This!\ results under two training conditions. All metrics are expressed as percentages.}
\label{tab:beat_tracking}
\end{table}

\subsubsection{Performance MIDI-to-Score Transcription}
\label{sec:performance_to_score}

We evaluate PM2S~\cite{liu2022performance} on the prediction of score-related attributes from performance MIDI, including beat and downbeat labels, hand assignment, key and time signatures, and rhythmic quantisation. We report positive-class F1 for beat and downbeat labels, macro-F1 for hand assignment and key/time signatures, and support-weighted F1 for quantised onsets and note values. Here, beat and downbeat scores measure note-level label prediction rather than the audio-based event detection evaluated in the preceding task.

\autoref{tab:pm2s_results} shows that adding Gangqinpu improves most evaluated attributes on the Gangqinpu test set. Beat and downbeat F1 increase by 14.02 and 20.95 percentage points, respectively, while onset and note-value weighted F1 improve by 18.87 and 19.32 points. Hand assignment also improves, whereas time-signature macro-F1 decreases. Transfer to ASAP is mixed: beat, downbeat, and note-value prediction improve, while hand assignment, key/time signatures, and onset quantisation decline. These results demonstrate benefits for transcription within the Gangqinpu domain, but do not indicate uniform improvements across attributes or datasets.

\begin{table*}[t]
\centering
\small
\setlength{\tabcolsep}{6pt}
\begin{tabular}{@{}llrrrr@{}}
\toprule
& & \multicolumn{2}{c}{\textbf{ASAP test}} & \multicolumn{2}{c}{\textbf{Gangqinpu test}} \\
\cmidrule(lr){3-4}
\cmidrule(lr){5-6}
\textbf{Prediction target} & \textbf{Metric} $\uparrow$ & \textbf{ASAP} & \textbf{ASAP + G} & \textbf{ASAP} & \textbf{ASAP + G} \\
\midrule
Beat label       & F1          & 62.97 & \textbf{64.89} & 69.28 & \textbf{83.30} \\
Downbeat label   & F1          & 43.36 & \textbf{48.68} & 42.02 & \textbf{62.97} \\
Hand assignment  & Macro-F1    & \textbf{86.65} & 84.81 & 86.99 & \textbf{93.99} \\
Key signature    & Macro-F1    & \textbf{36.26} & 30.87 & 78.39 & \textbf{79.33} \\
Time signature   & Macro-F1    & \textbf{45.40} & 43.96 & \textbf{51.00} & 45.43 \\
Quantised onset  & Weighted-F1 & \textbf{53.59} & 49.71 & 55.67 & \textbf{74.54} \\
Note value       & Weighted-F1 & 40.39 & \textbf{41.37} & 34.86 & \textbf{54.18} \\
\bottomrule
\end{tabular}
\caption{PM2S results under ASAP-only and combined training (ASAP + G, where G denotes Gangqinpu). All scores are percentages. Bold indicates the better training condition for each target within each test set.}
\label{tab:pm2s_results}
\end{table*}

\subsubsection{Expressive Performance Rendering}
\label{sec:expressive_rendering}

We evaluate the adapted S2A~\cite{tang2025towards} model to predict expressive performance attributes from musical scores. We report mean absolute error (MAE) for MIDI velocity, inter-onset interval (IOI), and note duration, with timing errors expressed in milliseconds. These metrics measure agreement with the reference performance rather than perceptual expressiveness.

As shown in \autoref{tab:s2a_results}, adding Gangqinpu reduces all three errors on the Gangqinpu test set. Velocity MAE decreases from 17.50 to 13.05, IOI MAE from 78.51 to 54.66~ms, and duration MAE from 309.47 to 269.04~ms. On ASAP, all three errors increase modestly, indicating that the benefits do not transfer uniformly across domains. Together, these results support the utility of Gangqinpu for modeling performances within its domain, while highlighting the challenge of improving performance across different musical and data distributions.

\begin{table}[t]
\centering
\small
\setlength{\tabcolsep}{4pt}
\begin{tabular}{@{}lccc@{}}
\toprule
\textbf{Training data} & \textbf{Velocity} $\downarrow$ & \textbf{IOI (ms)} $\downarrow$ & \textbf{Duration (ms)} $\downarrow$ \\
\midrule
\multicolumn{4}{c}{\textit{ASAP test}} \\
\midrule
ASAP & \textbf{16.40} & \textbf{66.40} & \textbf{999.66} \\
ASAP + Gangqinpu & 16.62 & 68.96 & 1,058.28 \\
\midrule
\multicolumn{4}{c}{\textit{Gangqinpu test}} \\
\midrule
ASAP & 17.50 & 78.51 & 309.47 \\
ASAP + Gangqinpu & \textbf{13.05} & \textbf{54.66} & \textbf{269.04} \\
\bottomrule
\end{tabular}
\caption{Expressive performance rendering results using the adapted S2A model. All metrics are MAE. Velocity errors are in MIDI velocity units.}
\label{tab:s2a_results}
\end{table}

\section{Conclusion}
\label{sec:conclusion}

We introduced Gangqinpu, a multi-genre piano dataset comprising 18,786 paired performances and 1,007.1 hours of source audio, with transcribed performance MIDI and musical scores connected by automatic note-level alignments. Our construction pipeline combines expressive performance filtering, confidence-aware beat annotation, genre labeling, and leakage-aware splitting to support multiple downstream tasks. Experiments demonstrate improved beat tracking on both Gangqinpu and ASAP, alongside gains in most performance MIDI-to-score transcription targets and all evaluated rendering metrics on Gangqinpu. Transfer to ASAP remains mixed for the latter two tasks, highlighting the challenges of cross-domain generalization. Future work will focus on manually auditing automatic annotations.

\bibliography{ref}

%
%
%
%

\end{document}